\RequirePackage{amsmath}
\documentclass[runningheads]{llncs}
\usepackage[T1]{fontenc}
\usepackage{graphicx}
\usepackage{tikz}
\usetikzlibrary{quantikz2}
\usepackage{booktabs}
\usepackage{makecell}
\usepackage{multirow}
\usepackage{subfig}
\usepackage{tabularx}
\usepackage{siunitx}
\usepackage{adjustbox}

\usepackage[capitalise]{cleveref}
\creflabelformat{equation}{#2#1#3}

\begin{document}
\title{Evaluating Parameter-Based Training Performance of Neural Networks and Variational Quantum Circuits}
\titlerunning{Parameter-Based Training Performance of NNs and VQCs}
%
\author{
Michael Kölle \and
Alexander Feist \and
Jonas Stein \and
Sebastian Wölckert \and
Claudia Linnhoff-Popien
}

\authorrunning{M. Kölle et al.}
%
\institute{LMU Munich, Oettingenstraße 67, Germany \\
\email{michael.koelle@ifi.lmu.de}}
\maketitle              
\begin{abstract}
In recent years, neural networks (NNs) have driven significant advances in machine learning. However, as tasks grow more complex, NNs often require large numbers of trainable parameters, which increases computational and energy demands. Variational quantum circuits (VQCs) offer a promising alternative: they leverage quantum mechanics to capture intricate relationships and typically need fewer parameters. In this work, we evaluate NNs and VQCs on simple supervised and reinforcement learning tasks, examining models with different parameter sizes. We simulate VQCs and execute selected parts of the training process on real quantum hardware to approximate actual training times. Our results show that VQCs can match NNs in performance while using significantly fewer parameters, despite longer training durations. As quantum technology and algorithms advance, and VQC architectures improve, we posit that VQCs could become advantageous for certain machine learning tasks.

\end{abstract}

\section{Introduction}\label{sec:introduction}
Machine learning has advanced rapidly in recent years, with neural networks (NNs) playing a pivotal role in this progress \cite{alom2019state}. NNs have propelled significant breakthroughs in areas such as image recognition \cite{krizhevsky2012imagenet}, natural language processing \cite{vaswani2017attention}, and game-playing \cite{silver2016mastering}. However, as tasks grow more complex, NNs often require a large number of trainable parameters, which increases computational and energy demands \cite{strubell2020energy,brown2020language}.

Variational quantum circuits (VQCs) represent a promising alternative to classical NNs \cite{du2020expressive,cerezo2021variational}. They harness quantum mechanics to model intricate relationships and usually need fewer parameters \cite{schuld2020circuit,lockwood2020reinforcement}. Despite being in an early stage, quantum computing is advancing quickly, and noisy intermediate-scale quantum (NISQ) devices are already available. These devices enable researchers to explore and benchmark quantum algorithms in realistic conditions. VQCs are well-suited to NISQ hardware since they tolerate the noise levels inherent in these devices \cite{fontana2021evaluating,cerezo2021variational}.

In this work, we evaluate the potential of VQCs relative to NNs on simple supervised and reinforcement learning tasks. We compare models with varying parameter counts to identify where VQCs may be advantageous. We carry out most VQC experiments on a simulator but approximate real hardware training times by running selected circuits on actual quantum devices. Our findings align with prior work, showing that VQCs can achieve performance similar to NNs while using fewer parameters. Although training VQCs takes longer, we suggest that continued advances in quantum technology, improvements in VQC architectures, and algorithmic optimizations may make VQCs appealing for certain applications. All code for the experiments is available here\footnote{https://github.com/alexander-feist/nn-vqc-params}. 

In \cref{sec:related-work} we offer context and examine related work. \cref{sec:approach} details our approach, focusing on the architectures of NNs and VQCs. \cref{sec:setup} outlines the experimental configuration, describing both supervised and reinforcement learning tasks. Finally, we present and discuss our results in \cref{sec:results}.

\section{Related Work}\label{sec:related-work}
This section presents related studies on quantum supervised learning in \cref{sec:rw-qsl} and quantum reinforcement learning in \cref{sec:rw-qrl}. To our knowledge, no existing work thoroughly compares NNs and VQCs for machine learning tasks with a detailed focus on model architectures, parameter counts, and training times.

\subsection{Quantum Supervised Learning}\label{sec:rw-qsl}
Quantum computing shows considerable promise in supervised learning (SL), particularly through hybrid quantum-classical approaches that combine VQCs with classical optimization \cite{schuld2019quantum,schuld2020circuit,mitarai2018quantum,havlivcek2019supervised}. Schuld and Killoran \cite{schuld2019quantum} proposed two classification methods that embed classical data into high-dimensional quantum space. One approach uses VQCs, which capture complex relationships in classical datasets. Schuld et al. \cite{schuld2020circuit} introduced a scalable VQC architecture and demonstrated, via simulation, that it achieves strong SL performance with fewer trainable parameters than classical NNs. Their design inspires the architecture used in our work. We compare models with varying parameter counts and approximate the time required to train on real quantum hardware.

Similarly, Mitarai et al. \cite{mitarai2018quantum} proposed \emph{quantum circuit learning}, which employs low-depth VQCs and classical optimization to approximate nonlinear functions in SL tasks. They discussed a potential quantum advantage for high-dimensional classification.

\subsection{Quantum Reinforcement Learning}\label{sec:rw-qrl}
Chen et al. \cite{chen2020variational} illustrated that VQCs can perform well in reinforcement learning (RL) by approximating the action-value function through Q-learning in simple discrete environments. Inspired by this work, we use their custom Frozen Lake environment and Q-learning approach to evaluate multiple VQCs and NNs with varying parameter counts.

Lockwood et al. \cite{lockwood2020reinforcement} extended this idea to the CartPole and Blackjack environments \cite{brockman2016openaigym}, which feature continuous state spaces. They showed that VQCs can match the performance of classical NNs while using fewer parameters. Kruse et al. \cite{kruse2023variational} explored architectural factors in VQCs for the Pendulum and LunarLander tasks \cite{brockman2016openaigym}, revealing that design choices such as input encoding, layering, and qubit count strongly affect outcomes. Although their VQCs employed about 96\% fewer parameters than the NNs, the NNs achieved higher rewards, and the VQCs faced challenges with scalability and robustness. Their study relied on proximal policy optimization and simulations rather than actual hardware.

Kölle et al. \cite{kolle2023multi} proposed a multi-agent quantum RL approach with evolutionary optimization to mitigate barren plateaus \cite{mcclean2018barren}, again showing that VQCs can match NN performance while using over 97\% fewer parameters. In contrast, our work focuses on a simple single-agent RL scenario for evaluating NNs and VQCs. Results such as those by Kölle et al.\ suggest that multi-agent setups may further accentuate the parameter savings of VQCs over NNs.

\section{Approach}\label{sec:approach}
In this work, we compare the training performance of classical NNs to VQCs on selected machine learning tasks. For each task, we evaluate a set of NNs with varying parameter counts and a corresponding set of VQCs that also differ in their parameter counts, aiming to identify one NN and one VQC with comparable performance. To ensure fair comparison, both models function as black-box components within the same classical learning algorithm. Although we conduct the VQC experiments primarily with a quantum simulator, we approximate the training times on real quantum hardware by running selected circuits—collected from simulator-based training—on an actual device. By comparing the simulator’s execution durations to those on the real machine, we estimate the training times on current quantum hardware. 

\cref{sec:nn-architecture} describes the fully connected feedforward NN, while \cref{sec:vqc-architecture} details the proposed VQC, including its data-encoding methods, variational layers, and measurement strategy.

\subsection{Classical Neural Network Architecture}\label{sec:nn-architecture}
We employ a fully connected feedforward NN. The input layer has as many nodes as the dimensionality of the input, followed by one or more hidden layers whose quantities and sizes are hyperparameters. Each hidden layer applies a rectified linear unit (ReLU) activation element-wise \cite{krizhevsky2012imagenet,nair2010rectified}:
\begin{equation}\label{eq:relu}
\text{ReLU}(z) = \max(0,z),
\end{equation}
where \(z\) is the pre-activation value of each node. The output layer has as many nodes as the number of possible outputs and uses a softmax activation to produce a probability distribution:
\begin{equation}\label{eq:softmax}
\text{Softmax}(\mathbf{z})_i = \frac{\exp(z_i)}{\sum_{j=1}^{K}\exp(z_j)},
\end{equation}
where \(\text{Softmax}(\mathbf{z})_i\) is the probability of the \(i\)-th output, \(K\) is the number of outputs, and \(\mathbf{z} = [z_1, z_2, \dots, z_K]\) contains the output logits \cite{krizhevsky2012imagenet,bridle1990probabilistic,goodfellow2016deep}.

\subsection{Variational Quantum Circuit Architecture}\label{sec:vqc-architecture}
The proposed VQC follows a circuit centric design  \cite{schuld2020circuit} and involves three stages: state preparation, variational layers, and measurement.

\subsubsection{State Preparation}
Classical input data can be embedded via angle embedding or amplitude embedding  \cite{larose2020robust}, chosen as a hyperparameter.

\paragraph{Angle Embedding}
Angle embedding maps each input feature to a rotation angle, using at least as many qubits as input dimensions. We employ \(X\)-axis rotations, with input values scaled into \([0,\pi]\).

\paragraph{Amplitude Embedding}
Amplitude embedding directly maps input values to the amplitudes of an \(n\)-qubit state  \cite{schuld2017implementing,schuld2020circuit}, which requires \(\lceil \log_2(D) \rceil\) qubits to represent \(D\)-dimensional data. We normalize the padded input vector and use the state preparation technique of Mottonen et al.  \cite{mottonen2004transformation}.

\subsubsection{Variational Layers}
\begin{figure}[htb]
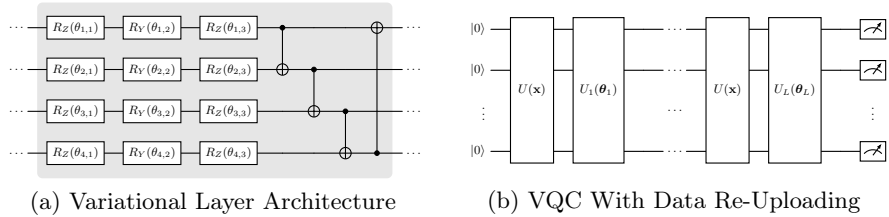

    \centering
    \subfloat[Variational Layer Architecture\label{fig:vqc-layer}]{
        \adjustbox{max width=0.48\linewidth}{\input{assets/vqc_layer.tex}}
    }
    \hfill
    \subfloat[VQC With Data Re-Uploading\label{fig:vqc-data-reup}]{
        \adjustbox{max width=0.48\linewidth}{\input{assets/vqc_data_reup.tex}}
    }
    \caption[Variational Quantum Circuit Designs]{(a) The \(l\)-th variational layer with 4 qubits, where \(\boldsymbol{\theta}_l = [\theta_{1,1}, \theta_{1,2}, \dots, \theta_{4,3}]\) are the trainable parameters for layer \(l\). (b) A VQC with \(L\) layers and repeated data embedding; \(U(\mathbf{x})\) encodes the input \(\mathbf{x}\), and \(U_l(\boldsymbol{\theta}_l)\) represents the trainable operations in layer \(l\) \cite{cerezo2021variational,schuld2020circuit}.}
    \label{fig:combined-vqc}
\end{figure}

Inspired by Schuld et al. \cite{schuld2020circuit}, we use variational layers comprising three single-qubit rotations (\(R_Z, R_Y, R_Z\)) per qubit, followed by CNOT gates for entanglement (\cref{fig:vqc-layer}). The trainable parameters \(\boldsymbol{\theta}\), initialized randomly in \([-1,1]\), are passed through 
\begin{equation}\label{eq:re-mapping}
\varphi(\mathbf{z}) = \pi \cdot \tanh(\mathbf{z}),
\end{equation}
to constrain angles to \((-\pi,\pi)\) \cite{kolle2022improving,kolle2023weight}. We also use data re-uploading \cite{perez2020data,skolik2022quantum}, which embeds the classical input values before every variational layer (\cref{fig:vqc-data-reup}).

\subsubsection{Measurement}
We measure the expectation value of the Pauli-\(Z\) operator on the first \(K\) qubits, where \(K\) matches the output dimension, then add a bias term to each measured value. These biases, initialized in \([-0.001,0.001]\), are trainable. We apply a softmax function, analogous to \cref{eq:softmax}, to derive output probabilities. Because the Pauli-\(Z\) expectation ranges from \([-1,1]\), we include a trainable scaling parameter (initialized at 1) to enhance the VQC’s effective output range \cite{skolik2022quantum}. This parameter is updated during training, promoting flexibility similar to that of classical NNs.

\section{Experimental Setup}\label{sec:setup}
This section outlines the setup for our experiments and details their implementation. We rely primarily on the PyTorch deep learning library \cite{paszke2019pytorch} and the PennyLane quantum machine learning framework \cite{bergholm2018pennylane}. For VQC experiments, we use PennyLane’s \verb|default.qubit| device for statevector simulation. All experiments run on a Linux cluster with Intel\textsuperscript{\textregistered}~Core\textsuperscript{TM}~i9-9900 processors, and we measure classical computation times by tracking the difference in \verb|time.perf_counter()| values.

We fix seeds to ensure reproducibility. Each experiment is repeated ten times using different seeds (0 to 9) to obtain more robust results. We present the average performance and 95\% confidence intervals, estimated by bootstrapping with 1000 resamples (seed = 0). For simplicity, tables show the mean \(\pm\) margin format, even though the intervals may not be perfectly symmetric.

We explore multiple NNs and VQCs with varying parameter counts for each machine learning task. To achieve this, we conduct an exhaustive grid search over model-based hyperparameters, considering all possible combinations. 
This process yields sets of NNs and VQCs that span a range of parameter counts, allowing us to identify pairs of models with comparable performance but different complexities. We ensure an equal number of NNs and VQCs in each grid search.


\subsection{Supervised Learning Experiments}\label{sec:sl-experiments}
All SL experiments address classification tasks, where the model predicts one of several classes based on given input features. Besides loss, we track accuracy as performance metrics.
We use three datasets of moderate complexity: Iris, Wine, and WDBC (Wisconsin Diagnostic Breast Cancer). 
For each dataset, we scale features to \([0,1]\) and split the data into 75\% training and 25\% testing, further dividing the test portion evenly into validation and test sets. This yields 112:19:19 splits for Iris, 133:22:23 for Wine, and 426:71:72 for WDBC.

\subsubsection{Training and Hyperparameters}\label{sec:sl-experiments-training}
We use the same classical training loop for all SL tasks, interchanging NNs or VQCs as the model. Each model outputs probabilities for each class. We train for 50 epochs with cross-entropy loss and Adam \cite{kingma2014adam} at a learning rate of \(0.01\), using a batch size of 8. After each epoch, we check performance on the validation set. We then select the model checkpoint with the highest validation accuracy and evaluate it on the test set.

For the Iris and Wine datasets, the NN grid search covers \{1,2,3\} hidden layers \(\times\)\{3,6,9,12\} nodes per layer, producing 12 NNs. For the VQC, we use angle or amplitude embedding and vary the number of variational layers from 1 to 6, also yielding 12 configurations. Because WDBC has 30 features, we only use amplitude embedding for VQCs to avoid 30-qubit circuits. We still vary the layer count from 1 to 6 (6 VQCs). To match, we search for NNs with \{1,2\} hidden layers \(\times\)\{3,6,9\} nodes, yielding 6 NNs. \cref{tab:sl-hyperparameters} summarizes the hyperparameters.

\begin{table}[htb]
    \centering
    \caption[Supervised Learning Hyperparameter Values]{Hyperparameter values for SL experiments on Iris, Wine, and WDBC. Curly braces \(\{\cdot\}\) indicate the exhaustive grid search sets. Amp stands for amplitude embedding, Ang for angle embedding.}
    \begin{tabularx}{\linewidth}{Xlccc}
\toprule
&
\thead[l]{\textbf{Hyperparameter}} &
\thead{\textbf{Iris}} &
\thead{\textbf{Wine}} &
\thead{\textbf{WDBC}} \\
\midrule
\multirow{3}{*}{\thead[l]{\textbf{NN \&}\\\textbf{VQC}}}
& Learning Rate      & 0.01                 & 0.01                 & 0.01                 \\
& Number of Epochs   & 50                   & 50                   & 50                   \\
& Batch Size         & 8                    & 8                    & 8                    \\
\midrule
\multirow{2}{*}{\thead{\textbf{NN}}} 
& Hidden Layers      & \{1, 2, 3\}          & \{1, 2, 3\}          & \{1, 2\}             \\
& Nodes per Layer    & \{3, 6, 9, 12\}      & \{3, 6, 9, 12\}      & \{3, 6, 9\}          \\
\midrule
\multirow{2}{*}{\thead{\textbf{VQC}}} 
& Encoding           & \{Amp, Ang\}         & \{Amp, Ang\}         & \{Amp\}              \\
& Variational Layers & \{1, 2, 3, 4, 5, 6\} & \{1, 2, 3, 4, 5, 6\} & \{1, 2, 3, 4, 5, 6\} \\
\bottomrule
\end{tabularx}

    \label{tab:sl-hyperparameters}
\end{table}

\subsection{Reinforcement Learning Experiments}\label{sec:rl-experiments}
Our RL experiments use Q-learning in a custom Frozen Lake environment, with the models (NN or VQC) approximating the action-value function \(Q\).
We primarily track reward to assess how well the agent learns. We use the test reward (averaged over 50 test episodes) as the main performance metric, along with the final-100-episode mean reward as an additional indicator. Reward curves in this work show the moving average over up to the last 50 episodes to smooth short-term variance. Training time is the total time used exclusively for training at each episode.

\subsubsection{Frozen Lake Environment}
We use the deterministic (non-slippery) Frozen Lake environment \cite{brockman2016openaigym} with custom rewards, following Chen et al. \cite{chen2020variational}. The 4\(\times\)4 grid contains safe (frozen) tiles and holes. The agent starts in the top-left and must reach the bottom-right goal. Each step yields \(-0.01\), reaching the goal yields \(+1.0\), and falling into a hole yields \(-0.2\). The shortest path has 6 steps, for a maximum reward of \(0.95\). We randomized the environment tiles depending on the seed to make it more challenging. 



\subsubsection{Training and Hyperparameters}\label{sec:rl-experiments-training}
We train for 500 episodes, each limited to 100 steps. The model observes a 4-dimensional binary encoding of the state (one of 16 tiles). Its output contains four action values. We use a policy model and a target model with identical architectures, updating the target model every 20 steps.
Actions are selected via an \(\epsilon\)-greedy strategy, with \(\epsilon\) initially \(1.0\). After each episode, we multiply \(\epsilon\) by \(0.99\) until it drops to \(0.01\). We also employ experience replay \cite{lin1992self} with a replay memory of size 1000. At each step, we sample a batch of 16 transitions for training. The policy model is updated via Adam \cite{kingma2014adam} at a learning rate of \(0.01\), using mean squared error loss and a discount factor \(\gamma=0.95\). After training, we evaluate over 50 test episodes without exploration.
The grid search for NNs uses \{1,2,3\} hidden layers \(\times\)\{3,6,9,12\} nodes (12 configurations). For VQCs, it varies embedding technique (amplitude or angle) and the number of layers from 1 to 6 (12 configurations). \cref{tab:rl-hyperparameters} shows the hyperparameters.

\begin{table}[htb]
    \centering
    \caption[Reinforcement Learning Hyperparameter Values]{Hyperparameter values for RL on the custom Frozen Lake environment. Curly braces \(\{\cdot\}\) show exhaustive grid search sets. Amp denotes amplitude embedding, Ang denotes angle embedding.}
    \begin{tabularx}{\linewidth}{Xlc}
\toprule
&
\thead[l]{\textbf{Hyperparameter}} &
\thead{\textbf{Frozen Lake}} \\
\midrule
\multirow{10}{*}{\thead[l]{\textbf{NN \&} \textbf{VQC}}}
& Learning Rate            & 0.01                 \\
& Discount Factor $\gamma$ & 0.95                 \\
& Replay Memory Capacity   & 1000                 \\
& Batch Size               & 16                   \\
& Number of Episodes       & 500                  \\
& Max. Steps per Episode   & 100                  \\
& Initial $\epsilon$       & 1.0                  \\
& Decay Rate $\epsilon$    & 0.99                 \\
& Min. $\epsilon$          & 0.01                 \\
& Target Model Update      & Every 20 steps       \\
\midrule
\multirow{2}{*}{\thead{\textbf{NN}}} 
& Hidden Layers            & \{1, 2, 3\}          \\
& Nodes per Layer          & \{3, 6, 9, 12\}      \\
\midrule
\multirow{2}{*}{\thead{\textbf{VQC}}}
& Encoding                 & \{Amp, Ang\}         \\
& Variational Layers       & \{1, 2, 3, 4, 5, 6\} \\
\bottomrule
\end{tabularx}

    \label{tab:rl-hyperparameters}
\end{table}

\subsection{Executing Quantum Circuits on Real Quantum Hardware}\label{sec:setup-real-hardware}
We simulate VQCs with PennyLane but estimate real-hardware training times through IBM’s cloud-based Qiskit Runtime. Running full training on real hardware is costly, so we log certain circuits (inputs and parameters) during simulator-based training and re-run only those circuits on actual quantum processors. Specifically, we pick circuits from five epochs/episodes under seed 0 and unparameterize them with the logged values. We then convert the PennyLane circuits to OpenQASM~2 \cite{cross2017open} and import them into Qiskit \cite{javadi2024quantum}, executing on \verb|ibm_fez| (version 2) backed by IBM’s Heron R2 processor. For 4-qubit circuits, \verb|ibm_sherbrooke| was slightly faster, but for 5-qubit circuits, \verb|ibm_fez| is significantly faster.
We use the usage metric from Qiskit Runtime to approximate quantum execution time. By comparing usage times on real hardware to simulator times for the same circuits, we derive an average ratio and apply it to the circuit execution times of simulator-based training to estimate real-hardware training duration. We do not analyze circuit outputs or noise effects, and we use 1024 shots for each circuit to keep conditions consistent.

\section{Results}\label{sec:results}
This section presents the outcomes of our experiments and discusses their implications. All reported metrics are averages over ten runs (seeds 0--9). We begin by examining the grid-search outcomes to select specific NNs and VQCs with comparable performance for each task. \cref{sec:sl-results} details the SL results, while \cref{sec:rl-results} covers the RL results. \cref{sec:real-hardware-results} presents our estimates of training times on real quantum hardware for the chosen VQCs. Finally, \cref{sec:overall-results} discusses the overall findings.

\subsection{Supervised Learning Results}\label{sec:sl-results}
We focus on models whose test accuracy and training curves suggest similar performance. We first identify a high test accuracy achievable by at least one NN and one VQC, then select a representative NN–VQC pair that meets this standard with relatively short training times and comparable accuracy/loss evolution.

\begin{figure}[htb]
    \centering
    \subfloat[Iris Dataset\label{fig:sl-iris-training}]{
        \includegraphics[width=0.31\linewidth]{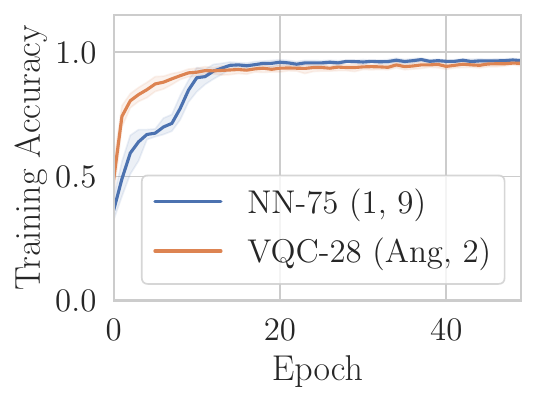}
    }
    \hfill
    \subfloat[Wine Dataset\label{fig:sl-wine-training}]{
        \includegraphics[width=0.31\linewidth]{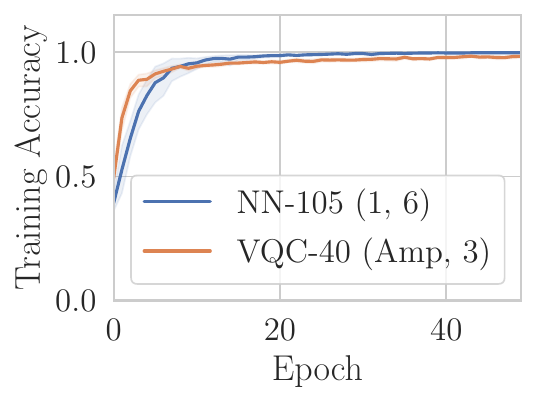}
    }
    \hfill
    \subfloat[WDBC Dataset\label{fig:sl-wdbc-training}]{
        \includegraphics[width=0.31\linewidth]{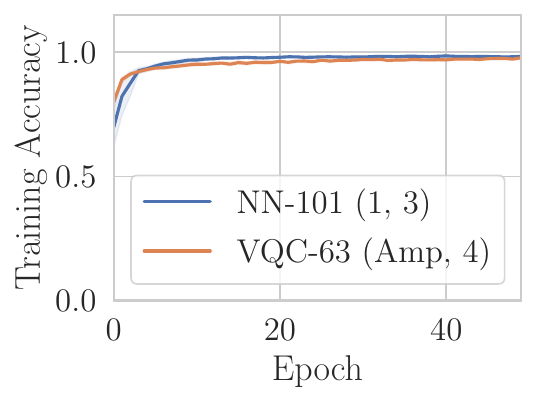}
    }
    \caption{Accuracy curves for each chosen NN and VQC. Averaged across ten runs (seeds 0--9); shaded areas are 95\% confidence intervals.}
    \label{fig:combined-sl}
\end{figure}

\subsubsection{Iris Dataset}\label{sec:sl-iris-results}
Our results show that the highest VQC test accuracy is 0.989, compared to 0.968 for the best NN. We consider models with at least 0.96 test accuracy as well-performing, yielding 4 of 12 NNs and 5 of 12 VQCs. Overall, VQCs slightly outperform NNs.
We select an NN with 75 parameters (1 hidden layer, 9 nodes) and a VQC with 28 parameters (angle embedding, 2 variational layers). They achieve comparable test accuracies (0.968 vs.\ 0.963) but require 1.8 seconds vs.\ 92.6 seconds of training, respectively. \cref{fig:sl-iris-training} shows that the VQC converges faster in accuracy early on but has a higher loss from about epoch 5 onward.


\subsubsection{Wine Dataset}\label{sec:sl-wine-results}
The best NN for the Wine dataset reaches 0.991 accuracy, and the best VQC reaches 0.974. We define 0.97 as our threshold for well-performing models, which is met by 8 of 12 NNs and 1 of 12 VQCs. NNs generally outperform VQCs here.
We pick an NN with 105 parameters (1 hidden layer, 6 nodes) and a VQC with 40 parameters (amplitude embedding, 3 variational layers). They reach 0.978 vs.\ 0.974 accuracy, requiring 1.7 seconds vs.\ 313.5 seconds. \cref{fig:sl-wine-training} shows that the VQC converges faster initially, but the NN ultimately has a lower loss.


\subsubsection{WDBC Dataset}\label{sec:sl-wdbc-results}
All tested models exceed 0.90 accuracy; the best NN reaches 0.972, while the best VQC attains 0.961. We set 0.96 as the threshold, met by 5 of 6 NNs and 1 of 6 VQCs. NNs again perform better overall.
We compare an NN with 101 parameters (1 hidden layer, 3 nodes) to a VQC with 63 parameters (amplitude embedding, 4 variational layers). Both reach 0.961 accuracy, requiring 5.4 seconds vs.\ 2482.9 seconds (about 41 minutes). \cref{fig:sl-wdbc-training} shows the VQC converges slightly faster initially, but the NN soon achieves a lower loss.


\subsection{Reinforcement Learning Results}\label{sec:rl-results}
\begin{figure}[htb]
    \centering
    \includegraphics[width=0.4\linewidth]{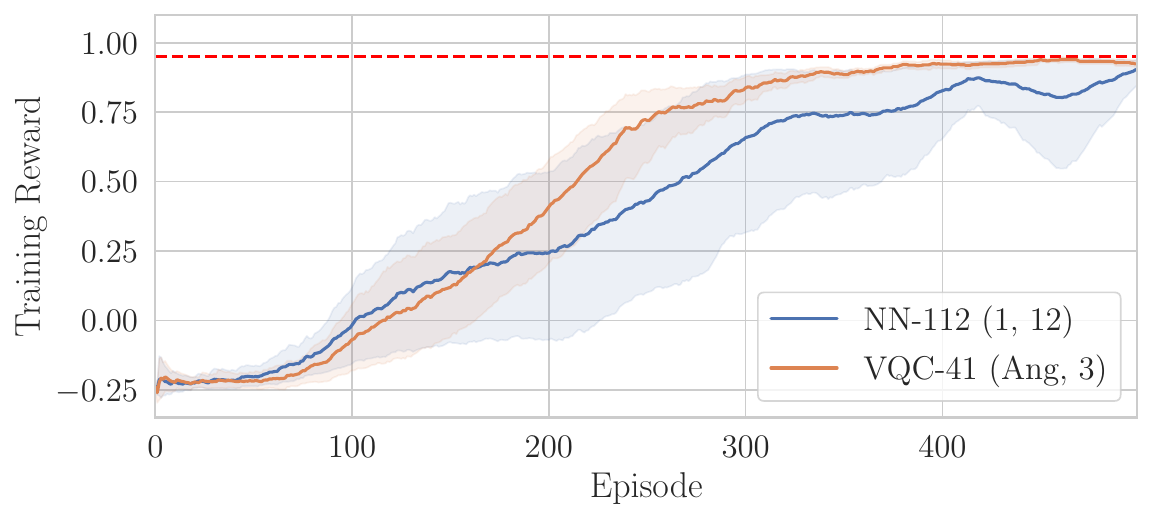}
    \caption[Reinforcement Learning: Reward During Training]{Training reward on Frozen Lake for the comparable NN (112 parameters) and VQC (41 parameters). Mean across ten runs (seeds 0--9); shaded areas are 95\% confidence intervals. The dashed red line (0.95) indicates the environment is solved.}
    \label{fig:rl-training}
\end{figure}

We use a custom deterministic Frozen Lake environment. We select models that achieve the maximum test reward of 0.95 and then pick an NN–VQC pair with comparable learning dynamics but relatively low training times.
2 of 12 NNs and 6 of 12 VQCs solve the environment (reward 0.95). VQCs generally outperform NNs here (see \cref{fig:rl-training}). However, training-time variance is high because episode length depends on agent behavior. We limit each episode to 100 steps, well above the optimal 6 steps, to allow exploration.

We select an NN with 112 parameters (1 hidden layer, 12 nodes) that requires 95.0 seconds and a VQC with 41 parameters (angle embedding, 3 variational layers) requiring 2511.4 seconds (about 42 minutes). Although an 85-parameter NN is slightly faster and eventually reaches a marginally higher average reward at the end of training, it is less stable, while the chosen NN is more comparable to the VQC in convergence. Our results show that the VQC converges faster and is more stable (narrower confidence interval).
\subsection{Training Times Using Real Quantum Hardware}\label{sec:real-hardware-results}
To estimate real-hardware training times for the chosen VQCs, we execute selected circuits on IBM’s Qiskit Runtime. We compare simulator-based and hardware-based execution times to compute a ratio, then apply this ratio to all simulator-based circuit calls. \cref{tab:real-hardware-time-per_circuit,tab:real-hardware-approximated-training-times} show the per-circuit times and final approximations. We focus on raw execution durations and do not factor in overhead or noise.

\begin{table}[htb]
    \centering
    \caption[Time Taken for Executing Circuits on Simulator and Real Hardware]{Mean execution time per circuit on the simulator vs.\ real hardware, plus their ratio. The VQC descriptions note the embedding technique and number of layers.}
    \begin{tabularx}{\linewidth}{
    Xlcc
    S[table-format=1.3, group-digits=false]
    S[table-format=1.3, group-digits=false]
    S[table-format=2.3, group-digits=false]
}
\toprule
\thead{\textbf{Task}} &
\thead{\textbf{VQC}} &
\thead{\textbf{Qubits}} &
\thead{\textbf{Circuit}\\\textbf{Depth}} &
\multicolumn{1}{c}{\thead{\textbf{Simulator}\\\textbf{(s)}}} &
\multicolumn{1}{c}{\thead{\textbf{Real Hard-}\\\textbf{ware (s)}}} &
\multicolumn{1}{c}{\thead{\textbf{Ratio}}} \\
\midrule
SL: Iris & VQC-28 (Ang, 2) & 4 &  17 & 0.011 & 0.314 & 28.995 \\
SL: Wine & VQC-40 (Amp, 3) & 4 & 100 & 0.034 & 0.349 & 10.295 \\
SL: WDBC & VQC-63 (Amp, 4) & 5 & 261 & 0.084 & 0.329 &  3.932 \\
RL       & VQC-41 (Ang, 3) & 4 &  25 & 0.016 & 0.322 & 20.406 \\
\bottomrule
\end{tabularx}

    \label{tab:real-hardware-time-per_circuit}
\end{table}

\begin{table}[htb]
    \centering
    \caption[Simulator-Based and Approximated Real Hardware Training Times]{Mean simulator-based training times vs.\ approximated real-hardware times, excluding overhead.}
    \begin{tabularx}{\linewidth}{
    Xlc
    S[table-format=4.1, group-digits=false]@{${}\,\pm\,{}$}S[table-format=3.1, group-digits=false]
    S[table-format=5.1, group-digits=false]@{${}\,\pm\,{}$}S[table-format=4.1, group-digits=false]
}
\toprule
\thead{\textbf{Task}} &
\thead{\textbf{VQC}} &
\thead{\textbf{Qubits}} &
\multicolumn{2}{c}{\thead{\textbf{Simulator (s)}}} &
\multicolumn{2}{c}{\thead{\textbf{Real Hardware (s)}}} \\
\midrule
SL: Iris & VQC-28 (Ang, 2) & 4 &   92.6 &   0.2 &  1806.1 &    4.1 \\
SL: Wine & VQC-40 (Amp, 3) & 4 &  313.5 &   1.3 &  2437.4 &   11.6 \\
SL: WDBC & VQC-63 (Amp, 4) & 5 & 2482.9 &  12.7 &  7732.5 &   37.2 \\
RL       & VQC-41 (Ang, 3) & 4 & 2511.4 & 219.3 & 39330.9 & 3458.8 \\
\bottomrule
\end{tabularx}

    \label{tab:real-hardware-approximated-training-times}
\end{table}

We observe that angle embedding simulations often have lower circuit depth than amplitude embedding, leading to smaller simulator runtimes but higher hardware-to-simulator time ratios. For 5-qubit circuits, the hardware ratio decreases, indicating that as qubit counts increase, the simulator grows slower relative to hardware, consistent with existing literature \cite{cicero2024simulation,chen201864}. Since overhead and noise are excluded, these estimates likely represent ideal scenarios, but further optimizations (e.g., fewer shots, specialized training environments) could significantly reduce real-hardware training times.

\subsection{Evaluating Training Performance}\label{sec:overall-results}
\begin{figure}[!t]
    \centering
    \includegraphics[width=0.6\linewidth]{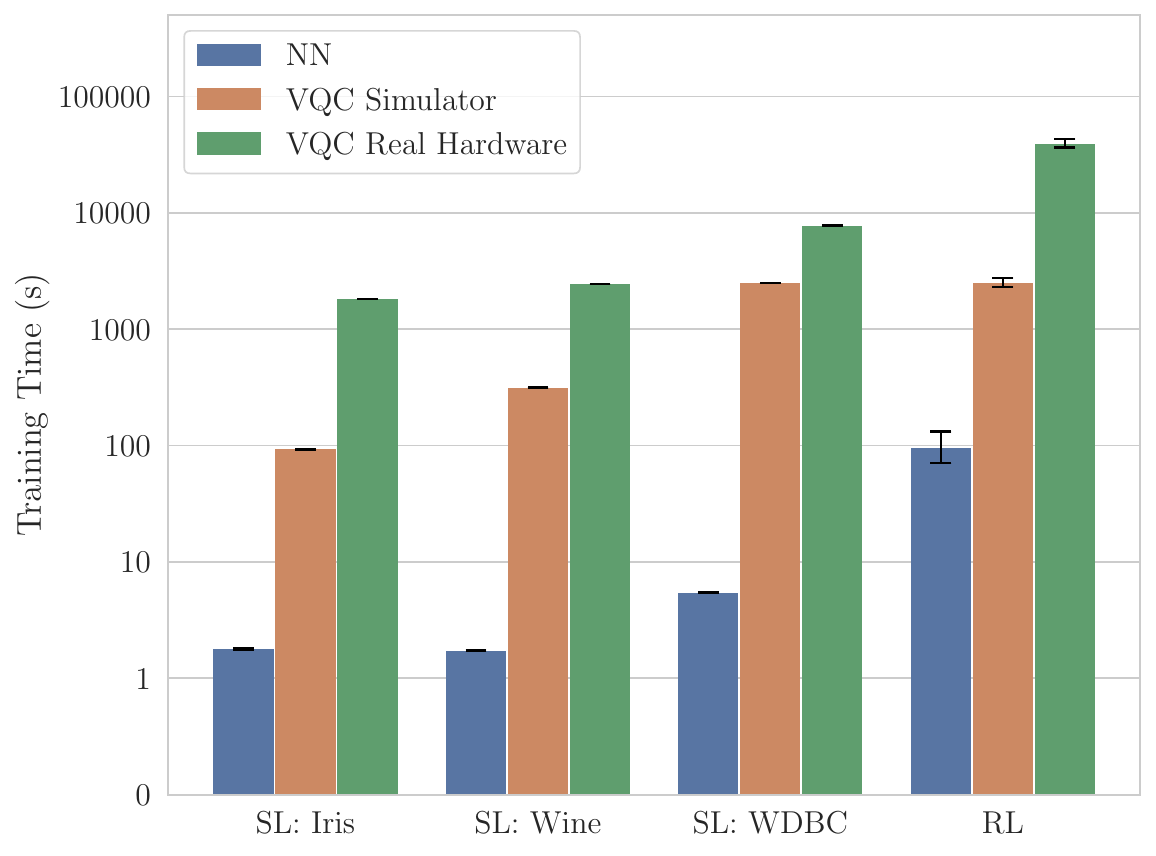}
    \caption[Training Times for Comparable Models]{Mean training times (seeds 0--9) for comparable NNs and VQCs. Error bars indicate 95\% confidence intervals. Note the logarithmic y-axis.}
    \label{fig:overall-training-time-comparison}
\end{figure}

\cref{tab:overall-comparison} and \cref{fig:overall-training-time-comparison} compare the chosen NNs and VQCs in terms of parameter count and training time. For similar performance:
\[
\text{VQCs require}\;
\begin{cases}
    62.7\%\;\text{fewer parameters (Iris SL)} \\
    61.9\%\;\text{fewer parameters (Wine SL)} \\
    37.6\%\;\text{fewer parameters (WDBC SL)} \\
    63.4\%\;\text{fewer parameters (Frozen Lake RL)}
\end{cases}
\]
but take far longer to train in our setup. SL tasks show tight confidence intervals for training time, whereas RL tasks exhibit broader uncertainty due to variable agent behavior.

\begin{table}[htb]
    \centering
    \caption[Training Times for Comparable Models]{Mean training times (with 95\% confidence intervals) for comparable NNs and VQCs.}
    \begin{tabularx}{\linewidth}{
    Xll
    S[table-format=2.1, group-digits=false]@{${}\,\pm\,{}$}S[table-format=2.1, group-digits=false]
    S[table-format=4.1, group-digits=false]@{${}\,\pm\,{}$}S[table-format=3.1, group-digits=false]
    S[table-format=5.1, group-digits=false]@{${}\,\pm\,{}$}S[table-format=4.1, group-digits=false]
}
\toprule
\multicolumn{1}{c}{} &
\multicolumn{2}{c}{\thead{\textbf{Model}}} &
\multicolumn{6}{c}{\thead{\textbf{Training Time (s)}}} \\
\cmidrule(lr){2-3} \cmidrule(lr){4-9}
\thead{\textbf{Task}} &
\thead{\textbf{NN}} &
\thead{\textbf{VQC}} &
\multicolumn{2}{c}{\thead{\textbf{NN}}} &
\multicolumn{2}{c}{\thead{\textbf{VQC \textsubscript{Simulator}}}} &
\multicolumn{2}{c}{\thead{\textbf{VQC \textsubscript{Real Hardware}}}} \\
\midrule
SL: Iris & NN-75  & VQC-28 &  1.8 &  0.0 &   92.6 &   0.2 &  1806.1 &    4.1 \\
SL: Wine & NN-105 & VQC-40 &  1.7 &  0.0 &  313.5 &   1.3 &  2437.4 &   11.6 \\
SL: WDBC & NN-101 & VQC-63 &  5.4 &  0.0 & 2482.9 &  12.7 &  7732.5 &   37.2 \\
RL       & NN-112 & VQC-41 & 95.0 & 30.7 & 2511.4 & 219.3 & 39330.9 & 3458.8 \\
\bottomrule
\end{tabularx}

    \label{tab:overall-comparison}
\end{table}

\cref{tab:ratio-comparison} compares parameter counts and training-time ratios. For the RL task, equalizing the two training times would require the VQC to be about 414 times faster, which may sound large but could become feasible as quantum hardware matures much faster than classical systems. Architectural and algorithmic improvements—such as specialized VQC optimizers—may also reduce this ratio. Furthermore, although VQCs often converge faster in accuracy or reward, our fixed training schedule does not exploit early convergence. 

\begin{table}[t!]
    \centering
    \caption[Parameter and Training Time Ratios for Comparable Models]{Ratios of parameter counts and mean training times for comparable NNs and VQCs.}
    \begin{tabular}{
    l
    S[table-format=1.3, group-digits=false]
    S[table-format=3.3, group-digits=false]
    S[table-format=4.3, group-digits=false]
}
\toprule
\thead{\textbf{Task}} & 
\multicolumn{1}{c}{\thead{\textbf{VQC to NN}\\\textbf{Parameter Ratio}}} & 
\multicolumn{1}{c}{\thead{\textbf{VQC \textsubscript{Simulator} to NN}\\\textbf{Training Time Ratio}}} & 
\multicolumn{1}{c}{\thead{\textbf{VQC \textsubscript{Real Hardware} to NN}\\\textbf{Training Time Ratio}}} \\
\midrule
SL: Iris & 0.373 &  51.558 & 1005.866 \\
SL: Wine & 0.381 & 181.683 & 1412.422 \\
SL: WDBC & 0.624 & 457.458 & 1424.665 \\
RL       & 0.366 &  26.435 &  413.997 \\
\bottomrule
\end{tabular}

    \label{tab:ratio-comparison}
\end{table}

Training time grows with parameter count, though architecture also matters (e.g., qubit count, circuit depth, or layer widths). Even in our small-scale tasks, we see a trend of longer training times for larger models, especially for VQCs. This trend may be more pronounced in complex tasks where standard NNs can have millions of parameters, potentially offering a more substantial advantage to VQCs that require fewer parameters \cite{kolle2023multi,lockwood2020reinforcement}. However, it remains unclear whether VQCs can scale effectively to complex tasks and still match NNs \cite{qian2022dilemma,kruse2023variational}. Rather than replacing NNs outright, VQCs may find value in scenarios where they offer distinct benefits—especially if quantum hardware, training algorithms, and circuit designs continue to improve.

\section{Conclusion}\label{sec:conclusion}
We created a unified environment to compare classical NNs and VQCs as interchangeable models for multiple machine learning tasks. An exhaustive grid search over model-based hyperparameters allowed us to identify similarly performing models with correspondingly few parameters, enabling a fair comparison of training times. Despite using fewer parameters, the VQCs performed on par with NNs across our experiments, particularly excelling in the RL task. However, the VQCs required substantially longer training durations, with simulator-based training being 26 to 457 times slower than the NNs.

By executing a subset of circuits on real quantum hardware, we approximated current hardware training times for the VQCs. Because these tasks used at most five qubits, simulations were relatively fast, resulting in real-hardware training times that were longer than those on the simulator. Our findings underscore the simplicity of the tasks, yet suggest that as quantum technology matures, VQC-friendly algorithms improve, and circuit architectures evolve, VQCs may offer advantages for specific applications.

Future work could assess more complex tasks to deepen our understanding of VQCs’ potential compared to NNs. Furthermore, our real-hardware tests only assessed circuit execution times and not final outputs, leaving device fidelity unexamined. Subsequent studies might incorporate circuit results to determine how many shots are needed for reliable predictions on noisy hardware. Noise-aware simulators could serve as an additional tool to evaluate VQC performance under realistic conditions.

\section*{Acknowledgements}
This work is part of the Munich Quantum Valley, which is supported by the Bavarian state government with funds from the Hightech Agenda Bayern Plus. This paper was partly funded by the German Federal Ministry of Education and Research through the funding program “quantum technologies — from basic research to market” (contract number: 13N16196).

%
%
\bibliographystyle{splncs04}
\bibliography{main}

\end{document}